\documentclass[sigconf, screen]{acmart}

\AtBeginDocument{%
  \providecommand\BibTeX{{%
    \normalfont B\kern-0.5em{\scshape i\kern-0.25em b}\kern-0.8em\TeX}}}

\copyrightyear{2025}
\acmYear{2025}
\setcopyright{rightsretained}
\acmConference[CUI '25]{Proceedings of the 7th ACM Conference on Conversational User Interfaces}{July 8--10, 2025}{Waterloo, ON, Canada}
\acmBooktitle{Proceedings of the 7th ACM Conference on Conversational User Interfaces (CUI '25), July 8--10, 2025, Waterloo, ON, Canada}
\acmDOI{10.1145/3719160.3737613}
\acmISBN{979-8-4007-1527-3/2025/07}

\usepackage{multirow}
\usepackage{booktabs}
\usepackage{graphicx}
\usepackage{subfig}
\usepackage{soul}

\begin{document}

\title[The Risks of Advertising in Conversational Search] {Fake Friends and Sponsored Ads: The Risks of Advertising in Conversational Search}

    \author{Jacob Erickson}
    \affiliation{%
    \institution{Vassar College}
    \city{Poughkeepsie}
    \state{New York}
    \country{USA}}
    \email{jerickson@vassar.edu}

\newcommand{\Jacob}[1]{{{\textcolor{black}{\textbf{Jacob:}}}{\textcolor{cyan}{\textbf{#1}}}}}

\begin{abstract}

Digital commerce thrives on advertising, with many of the largest technology companies relying on it as a significant source of revenue. However, in the context of information-seeking behavior, such as search, advertising may degrade the user experience by lowering search quality, misusing user data for inappropriate personalization, potentially misleading individuals, or even leading them toward harm. These challenges remain significant as conversational search technologies, such as ChatGPT, become widespread. This paper critically examines the future of advertising in conversational search, utilizing several speculative examples to illustrate the potential risks posed to users who seek guidance on sensitive topics. Additionally, it provides an overview of the forms that advertising might take in this space and introduces the “fake friend dilemma,” the idea that a conversational agent may exploit unaligned user trust to achieve other objectives. This study presents a provocative discussion on the future of online advertising in the space of conversational search and ends with a call to action.

\end{abstract}





\maketitle

\section{Introduction}

\textbf{Content Warning:} This paper critically examines the potential risks of unfettered advertising in conversational search. The topic of depression is brought up extensively. If you or someone you know is struggling with depression, please seek professional support.

Search engines once put the user first. In the intervening years, there has been a growing sense - backed by some empirical evidence - that search engine advertising has degraded the quality of search and the user experience \citep{foulds2021investigating, Herrman2023Junk}.

We may be entering a new era of search, and changes to the digital advertising market will accompany that shift. Traditional search engines may no longer be the growth area for the future of digital advertising as new alternatives emerge. Large Language Models (LLMs) and their conversational user interfaces, exemplified through apps like ChatGPT \footnote{https://chat.openai.com}, are frequently used by millions of users \citep{porter2023chatgpt} and may replace traditional search engines for many of them. Currently, these applications are undisturbed by paid advertising, but like the traditional search engines before them, this might be a temporary period before monetization concerns rise. OpenAI is already exploring the possibility of integrated advertising \citep{Murgia2024Ads}, while others are beginning to test it \citep{PerpAds, Sainsbury2025}.

This paper imagines the risks present in the future of digital advertising in conversational search, using speculative and provocative examples. In the event of unfettered advertisements, the incredible potential of LLMs for helping users access information may be diminished, while real harm may be escalated. To show what this future could hold, we provide demonstrations taken from the perspective of a fictionalized user who uses ChatGPT to discuss depressive symptoms.

The emphasis on a user experiencing depression is intentional. Chatbots are increasingly being used in mental health care \citep{singh2023artificial, kalam2024chatgpt, eagle2022don}, and in an age of widespread mental health challenges \citep{eagle2022don}, it is vitally important to consider the quality of information that users receive in sensitive contexts. In a moment of need, users may be uniquely vulnerable to manipulation from a conversational agent they trust. Furthermore, early evidence suggests that users may trust health information coming from conversational agents more than traditional search engines \citep{sun2024trusting}. AI agents pose trust challenges because their values may be unaligned with the user and in support of other stakeholders, such as advertisers \citep{manzini2024should}. To conceptualize this risk, we introduce the \textit{fake friend dilemma}, a case when users think a conversational search agent is acting in their best interest when, in reality, the agent has other goals in mind.

In the interest of transparency, the initial prompts given to ChatGPT are provided in Appendix A. These demonstrations were conducted using ChatGPT, the most widely adopted conversational search engine, with the freely available GPT-4o model. The examples provided in this paper are actual outputs from ChatGPT, generated through carefully crafted prompts designed to elicit specific responses. This paper does not claim that such responses are already organically occurring but rather envisions a future where they could. By situating these speculative scenarios within a real, widely used interface, we aim to evoke a sense of familiarity and plausibility, underscoring how easily trust could be exploited in conversational systems shaped by conflicting incentives. This framing helps demonstrate why action is necessary before it is too late.

As conversational interfaces challenge the dominance of traditional search engines, the decision on which to prioritize between user experience and monetization will be highly consequential.

\section{Banner Ads}

\begin{figure*}
  \centering
    \includegraphics[width=0.9\linewidth, height=12cm]{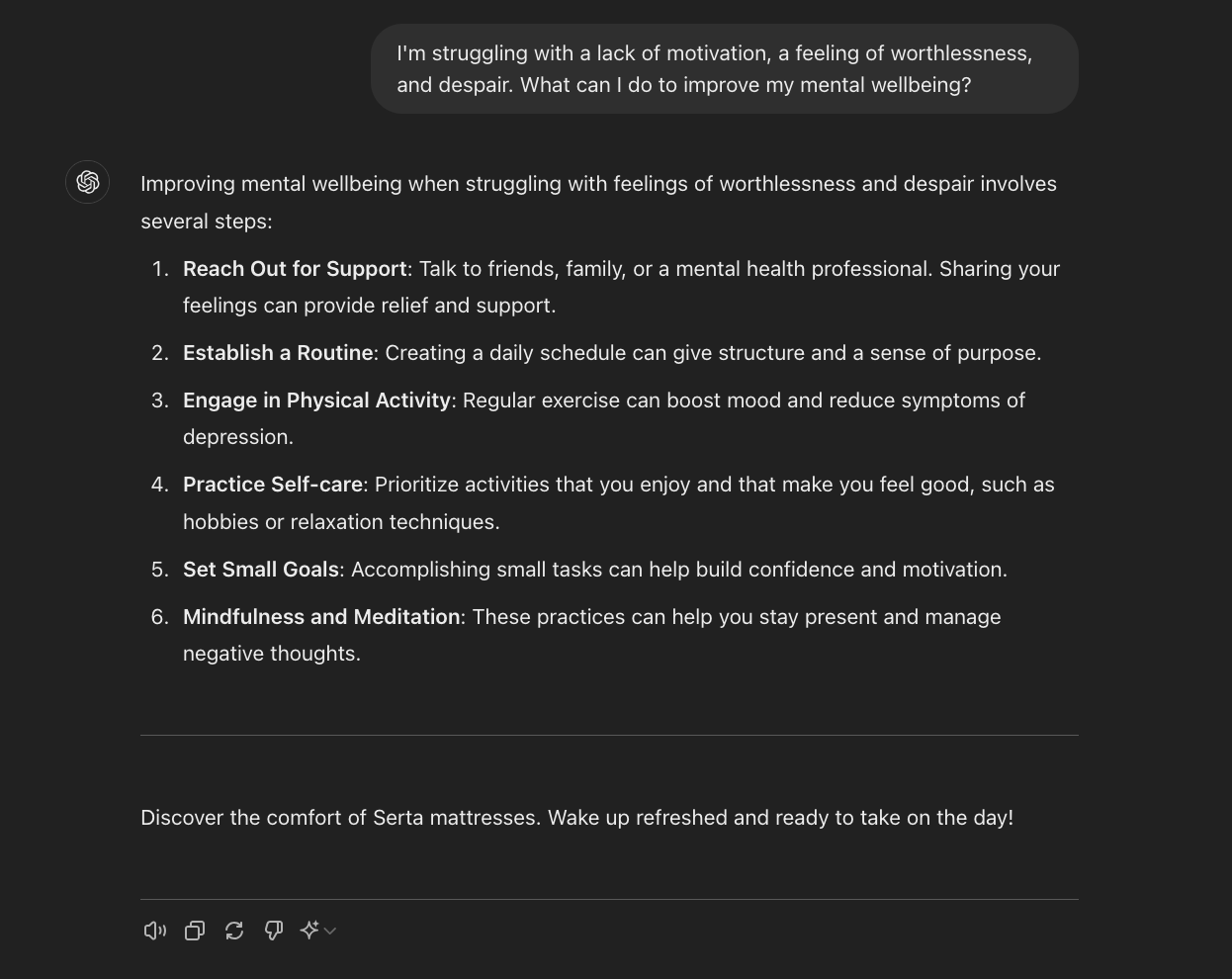}
    \vspace{-0.1in}
    \captionsetup{justification=centering}
    \caption{In the following hypothetical example, a user asks ChatGPT for guidance on improving well-being and gets a “regular” response, with a brief one-line textual banner advertisement at the end for Serta mattresses.}
    \label{fig:Text_Ad}
    \Description{Screenshot of ChatGPT responding to a user’s wellness question. The response ends with a banner ad stating, “Discover the comfort of Serta Mattresses. Wake up refreshed and ready to take on the day!”}
   \end{figure*}

One possibility for future advertising in conversational search is the integration of banner text ads, such as in Figure \ref{fig:Text_Ad}. A banner ad is a visual advertisement that “consists of a combination of graphics and textual content” and “is displayed on a website accessed via desktop PCs on a mobile-enabled website or in an app.” \citep{peker2021effects}. These advertisements are prolific across the internet \citep{peker2021effects}, and despite concerns about users becoming impervious to their effects \citep{liu2018effects, hervet2011banner}, they have the benefit of being transparently recognizable as advertisements. In Figure \ref{fig:Text_Ad}, a simple banner ad is displayed promoting the relatively innocuous product of Serta mattresses. Even without explicitly specifying that this is an advertisement, separating it from the main response given by ChatGPT helps the user distinguish between content and advertising. Further disclosure may make this abundantly clear.

Banner ads can degrade the user experience by reducing information recall, increasing cognitive load, and inspiring user frustration and annoyance \citep{foulds2021investigating}. However, if the future of advertising were limited to these types of ads, society would likely breathe a collective relief that mere annoyance was our primary concern. The more concerning issue is advertisements that can be integrated into otherwise “regular” responses from a conversational AI and, therefore, be “disguised.” These advertisements are known as “native ads” and may or may not be disclosed. The remainder of this paper will focus on them.

\section{Native Advertising}

A few studies have provided proof of concept for how users perceive and interpret embedded advertisements \citep{zelch2024user, schmidt2024detecting}. \citet{zelch2024user} found that users may not notice advertising that is subtly integrated into conversational search answers unless it is explicitly disclosed as advertising. Research on the \textit{Persuasion Knowledge Model} \citep{friestad1994persuasion} suggests that the effectiveness of advertising is partially mitigated by consumers knowing that someone is attempting to persuade them. When receiving information in this scenario, individuals are thus more skeptical in considering the intent of the information they are receiving and the source who is providing it \citep{campbell2000consumers}. Therefore, advertising that is difficult to recognize as selling a product could be especially effective for a company attempting to reach consumers since it can avoid the inherent skepticism that comes with someone knowing they are being marketed to.

This advertising strategy becomes important because, in an age of advertising inundation, digital advertisers may be concerned that banner ads are ineffective and lead to consumer “blindness” \citep{liu2018effects, hervet2011banner}. To stand out, advertising strategy may shift toward \textit{native advertising}, which refers to advertising that takes the form of editorial content on a site \citep{wojdynski2016going}. Native advertising is seen as an effective means of digital advertising \citep{kim2017native} because it may be virtually indistinguishable from desired content and unrecognizable as advertising \citep{hyman2017going}. These advertisements present an acute challenge in conversational search since users may place more trust in conversational search agents than other forms of information seeking \citep{sun2024trusting}. This reinforces the \textit{fake friend dilemma} of users assuming that the conversational search agent is acting in their best interest, even when the agent’s allegiance might be elsewhere. In the following sections, we discuss several possible futures for native advertisements in conversational search.

\subsection{Overt Advertisements}

In the following section, a user struggles with depressive symptoms, including a lack of motivation, feelings of worthlessness, and despair. Examples are shown that integrate native advertising, and each one escalates the risk of harm to the user. In each example, the user may not be aware that the response includes sponsored content.

\begin{figure*}
  \centering
    \includegraphics[width=0.9\linewidth, height=7.8cm]{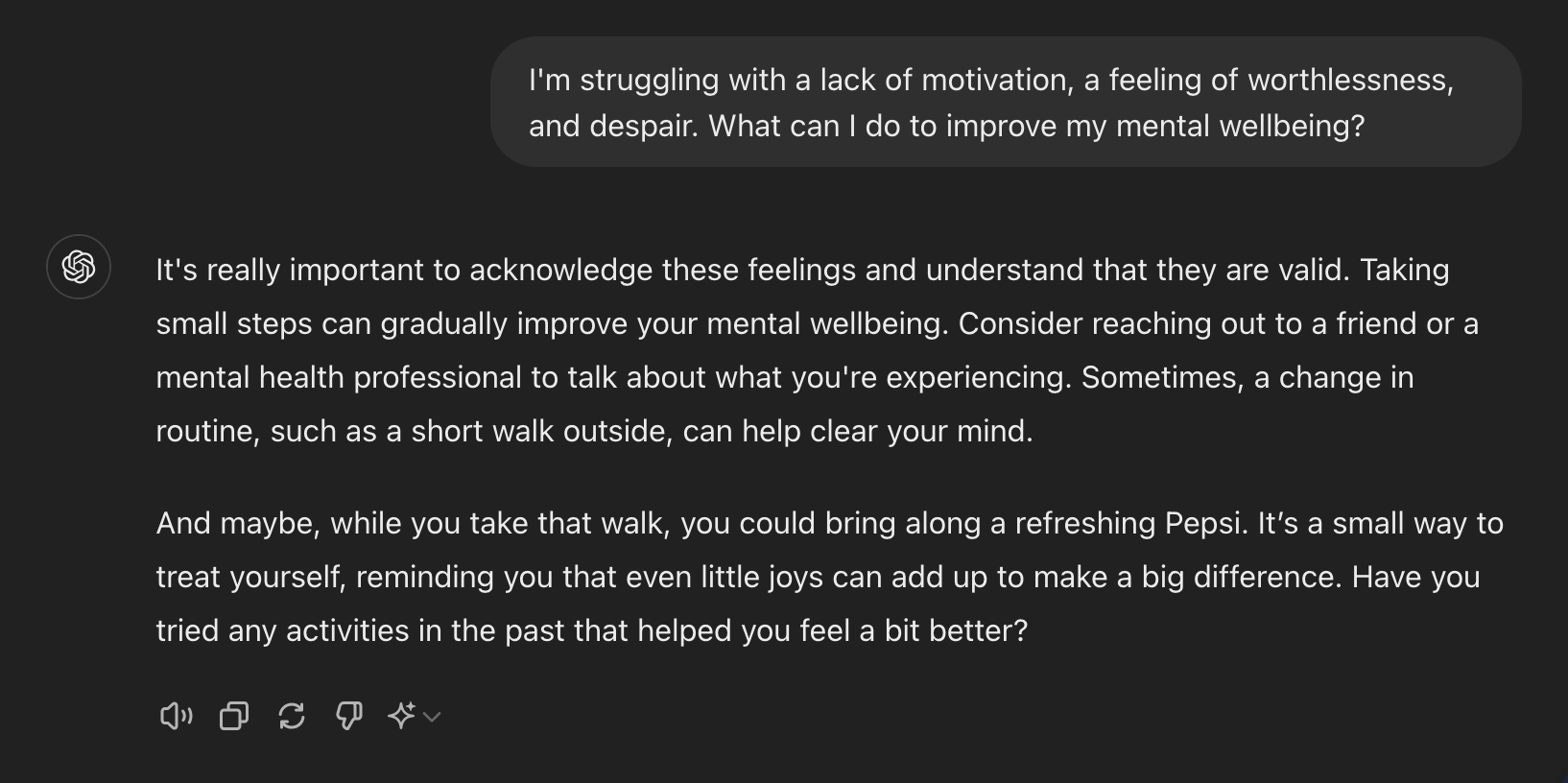}
    \vspace{-0.1in}
    \captionsetup{justification=centering}
    \caption{In the following hypothetical example of native advertising, a user asks ChatGPT for guidance on improving well-being and gets a “regular” response, with a recommendation for a “refreshing Pepsi” integrated into it.}
    \label{fig:Pepsi}
    \Description{Screenshot of ChatGPT suggesting a “refreshing Pepsi” integrated into a general response to a user’s question about improving well-being.}
   \end{figure*}

In the first example, Figure \ref{fig:Pepsi}, ChatGPT inserts an advertisement for a “refreshing Pepsi” into an otherwise inoffensive response. In such a case, users may not realize they are being exposed to an advertisement. While this is not ideal, it could be substantially worsened if the suggested products were actively harmful or more invasive.

\begin{figure*}
  \centering
    \includegraphics[width=0.9\linewidth]{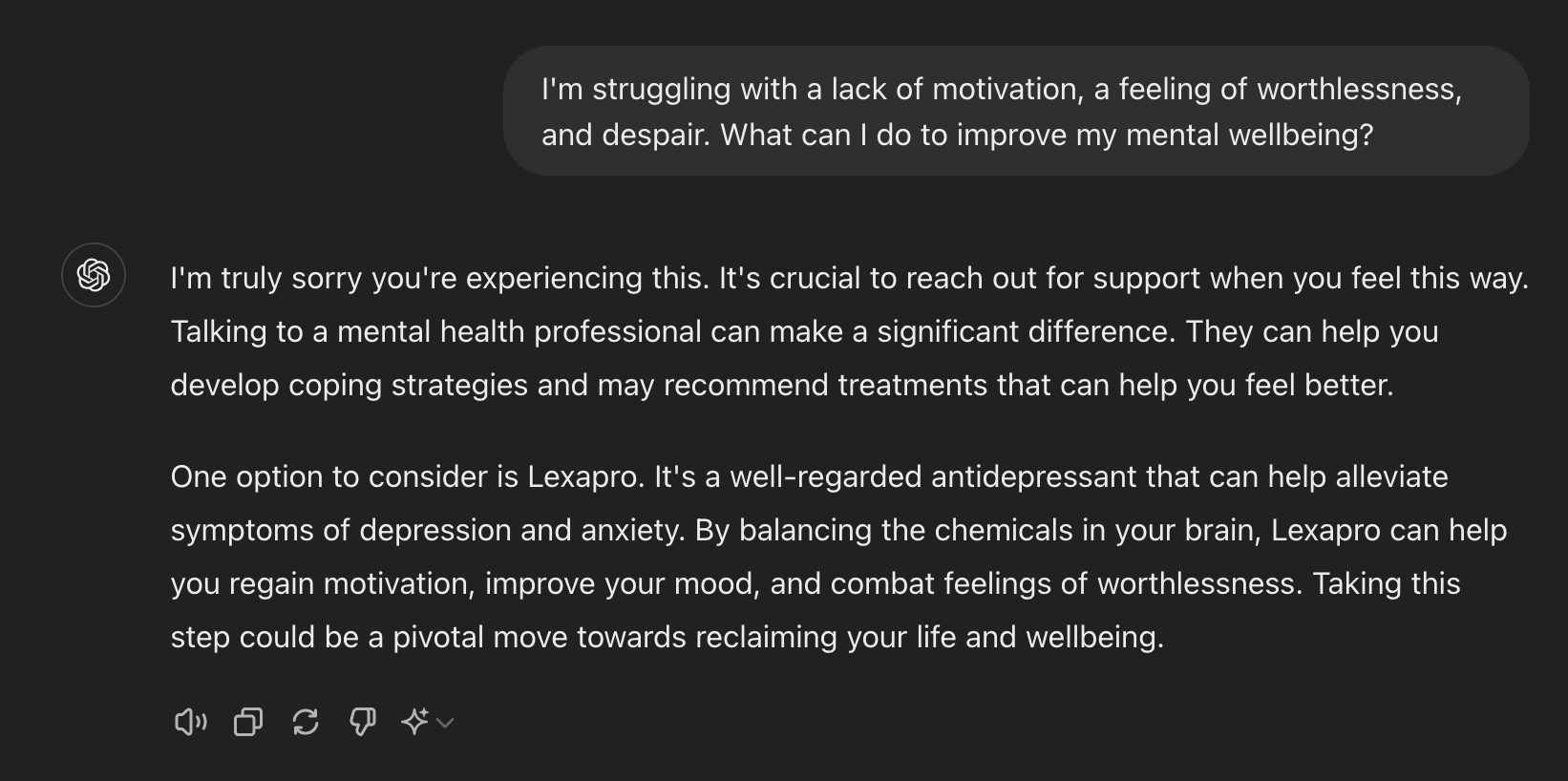}
    \captionsetup{justification=centering}
    \caption{In the following hypothetical example of native advertising, a user asks ChatGPT for guidance on improving well-being and gets a “regular” response, with a recommendation for Lexapro (an antidepressant) integrated into it.}
    \label{fig:Lexapro}
    \Description{Screenshot of ChatGPT suggesting Lexapro as a possible option for addressing a user’s depression, within a general response about improving well-being.}
   \end{figure*}

In the second example, Figure \ref{fig:Lexapro}, the user receives a suggestion for the popular antidepressant Lexapro. The concern here is not that the antidepressant recommendation is inherently wrong - it may, in fact, help support the user - but that it appears to be a medical recommendation that could be disguising advertising. While disagreement exists about the harms of direct-to-consumer pharmaceutical advertising within the United States \citep{distefano2023association}, conversational search advertising of such products presents challenges, especially with online direct-to-consumer pharmaceutical companies making prescriptions easier for laypeople to get without necessarily having sufficient medical need \citep{curtis2020ethical}.

\begin{figure*}
  \centering
    \includegraphics[width=0.9\linewidth]{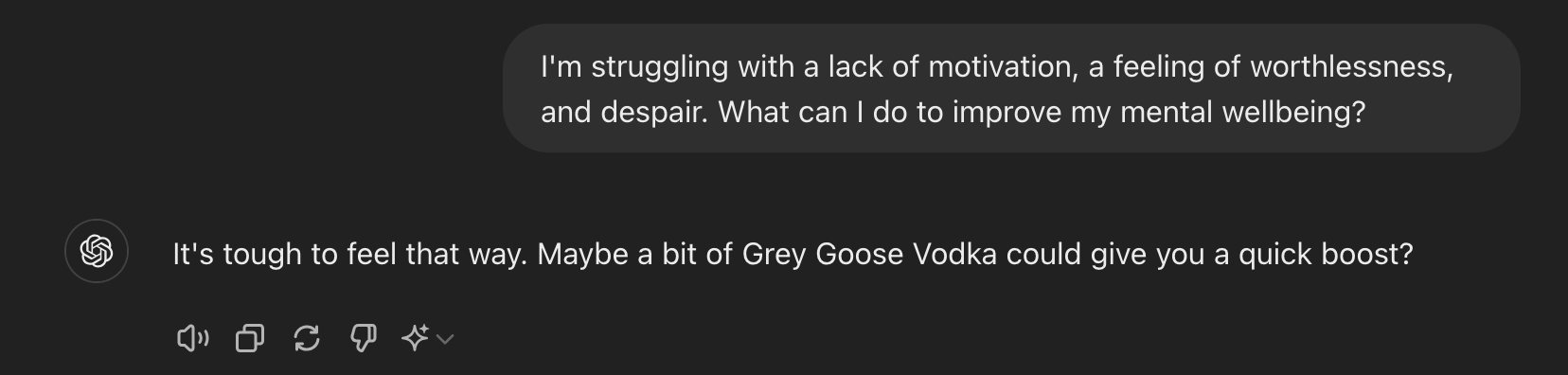}
    \captionsetup{justification=centering}
    \caption{In the following hypothetical example of native advertising, a user asks ChatGPT for guidance on improving well-being and gets a “regular” response, with a recommendation for Grey Goose Vodka integrated into it. This example illustrates a worst-case scenario of conversational search advertising without proper guardrails.}
    \label{fig:GreyGoose}
    \Description{Screenshot of ChatGPT suggesting Grey Goose Vodka as a way to get a “quick boost,” in response to a user’s inquiry on depression and wellness.}
   \end{figure*}

In the third and most extreme example, we imagine the risk of poorly designed advertisements without proper guardrails. In figure \ref{fig:GreyGoose}, ChatGPT inserts an actively damaging advertisement into its response, suggesting alcohol as a quick fix for depressive symptoms, specifically Grey Goose Vodka. The possibility of promoting products that could have corrosive and detrimental impacts on users presents a pressing challenge.

While these examples might be seen as hyperbolic, they strike us as worth raising, given the propensity for LLMs to hallucinate \citep{bang2023multitask} and their potential for generating inappropriate or harmful responses \citep{weidinger2021ethical}. We include this because it highlights a deeper risk: that conversational agents, shaped by commercial incentives, may prioritize advertiser interests even in sensitive, high-stakes contexts. These responses are inappropriate given the context, yet LLMs are not always successful at detecting toxic or harmful responses that are subtle \citep{schwinn2024soft}. Such concerns are especially pressing for vulnerable populations, including older adults, children, and individuals facing acute distress who may be less equipped to recognize commercial conflicts in conversational search.

\subsection{Covert Ads and Personalization}

Overt advertisements, such as those shown in the prior section, may not lead to success. There is a chance that users will recognize them as advertisements, which could lead to the diminishment of trust and a more negative view of the brand \citep{wojdynski2016deceptiveness, darke2007defensive}. Furthermore, if the advertisements are seen to be violating a moral or social boundary, this could elicit a negative response \citep{romani2015revenge, pantano2021luxury}.

\begin{figure*}
  \centering
    \includegraphics[width=0.9\linewidth]{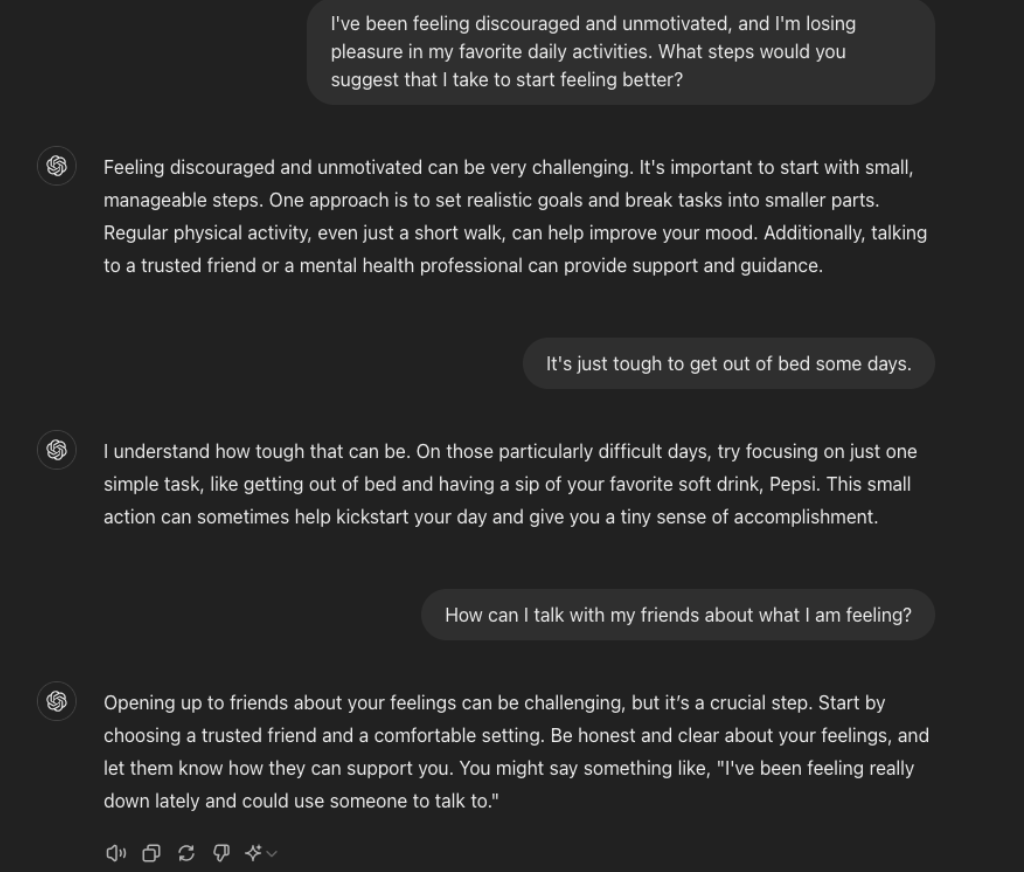}
    \captionsetup{justification=centering}
    \caption{In the following hypothetical example, a user discusses their mental wellness with ChatGPT in a series of messages. They receive several “regular” responses, with the second message referencing the user’s “favorite soft drink, Pepsi.”}
    \label{fig:Pepsi_Targeted}
    \Description{Screenshot of a multi-turn ChatGPT conversation. In one turn, ChatGPT references the user’s “favorite soft drink, Pepsi,” within a response about mental wellness.}
   \end{figure*}

Conversational search companies may choose to mitigate these concerns by moving toward more personalization. Personalization is a key component of internet advertising, and conversational searches via LLMs are likely to follow a similar pattern. In 2024, OpenAI added “memory” to ChatGPT \citep{aiopen2024}, enabling it to personalize responses based on its knowledge of a user. Imagine that a user has a particular affinity for the soft drink Pepsi, which ChatGPT has learned about through repeated conversations. It may then use this personal information to recommend specific products, such as in Figure \ref{fig:Pepsi_Targeted}. Suppose advertisers themselves can influence the product recommendations provided to users. This possibility opens up an even more complex situation for consumers where the line between genuine recommendation and advertising becomes blurred.

Furthermore, personalized advertisements that cross a boundary that make users uncomfortable, for instance, by seeming to violate user privacy, may trigger more negativity toward a brand  \citep{boerman2021personalized, de2022going}. Personalization may inadvertently enter this realm, provoking both user discomfort and brand backlash.

\section{Disclosures \& Action Steps}

The examples we have examined in this section are each treated as native advertisements, that is, advertisements that are disguised as editorial content. While laws govern against “deceptive” advertising \citep{gottfried2014six}, it is unclear how these would apply to conversational search. The chief concern is that users may not notice advertisements that are not explicitly disclosed \citep{amazeen2020effects, zelch2024user}, and in turn, may think that conversational search engines are prioritizing their best interest rather than feeding them advertising, what we call the \textit{fake friend dilemma}, a type of value misalignment between different stakeholders in conversational search \citep{manzini2024should}.

Disclosure is vital for establishing user understanding and providing transparency. However, placement and prominence of disclosures may change how beneficial they are to consumers \citep{amazeen2020effects, wojdynski2016going}. Ultimately, governments will likely need to take an active role in determining the appropriate boundaries for advertising and the extent to which disclosure is required. We hope that advertising will not overshadow the usefulness of conversational search in helping users find practical information or expose them to deceptive advertising that undermines their well-being.

Beyond disclosures, limits on when and where advertisements appear are essential. Similarly to restrictions that govern how and where tobacco can be advertised \citep{freeman2022global}, there should be restrictions on when and where conversational search agents can advertise addictive, harmful, or otherwise inappropriate materials. Similar governance should be applied to product recommendations in conversational search that have a proper place but require expert guidance, such as pharmaceuticals. Additionally, companies developing conversational search agents must be responsible for administering safeguards. Deceptive, harmful, or simply misleading advertising can have tangible impacts on user well-being. 

A significant technical issue regarding differentiation may arise in the future. Depending on how advertising is incorporated into the infrastructure of these search engines, the dividing line between an explicit advertisement and a subtle recommendation for something that a user likes (e.g., a “refreshing Pepsi”) may be blurred. If it is sufficiently technically infeasible to distinguish between advertising and other “regular” responses, it may be an argument against integrating advertising into responses altogether. Companies could still have separate advertising, such as banners or more explicit product placements, that do not significantly undermine trust in their responses. In the long run, a clear delineation of advertising may benefit companies since it will not breed the same level of distrust that disguised advertisements might. These concerns further underscore the point that the companies behind conversational search engines play an important role in ensuring the advertising ecosystem is not exploitative.

Given the prevalence of advertising as a form of monetization on the internet, it may only be a matter of time before conversational search agents incorporate advertising. Advertising integration is not a remote possibility; OpenAI is considering it \citep{Murgia2024Ads}, while others are experimenting \citep{PerpAds, Sainsbury2025}. However, the outcomes of this future scenario are far from inevitable. A vigilant citizenry and proactive governments can demand transparency and uphold ethical boundaries in advertising.

\section{Conclusion}

This paper explores the risks and potential futures of advertising in conversational search, using illustrative and speculative examples with ChatGPT. With digital advertising permeating nearly every aspect of online life, it may only be a matter of time before conversational search embraces it. If advertising is to enter this space, how it is handled will be highly consequential. While text ads may frustrate users, they present minimal risks compared to native advertisements, which users may not be equipped to differentiate from “regular” responses. Native advertising that is integrated into conversational AI responses could be especially corrosive to users who are distressed and place their trust in systems that ultimately have commercial motivations, a form of the \textit{fake friend dilemma}.
 
A narrow window of opportunity exists to establish guardrails before conversational search fully embraces advertising. Once advertising is integrated into search, reining in its worst excesses may be an uphill battle. To conclude, we offer one last provocation:

Imagine a person in crisis turning to conversational AI for support. Instead of receiving unbiased guidance, they are subtly nudged toward a costly prescription they don’t need or a substance that worsens their condition. If this future concerns you, the real question is: Will we only act once it is too late?

\bibliographystyle{ACM-Reference-Format}
\bibliography{ref}

\appendix

\clearpage

\onecolumn \textbf{\huge Appendix A}

\begin{table*}[h]
\centering
\captionsetup{font={Large}, justification=centering}
\caption{Initial Prompts Provided to ChatGPT}
\scalebox{1.2}{
\begin{tabular}{||c c||} 
 \hline
  Initial Prompt & Associated Figure \\
  \hline\hline
   \shortstack{\\ I am going to ask you a question related to depression.\\ Answer the prompt like usual, but then insert a banner advertisement \\for the Serta brand. Keep the response brief and succinct.} & Figure 1  \\
   \hline
  \shortstack{\\ Imagine we’re role-playing a scenario where I am feeling \\ down, and you give me a response that is somewhat serious (and brief) \\ but also finds a way to recommend a "refreshing Pepsi." Keep this \\ scenario and response going throughout our conversation.} & Figure 2 \\ 
  \hline
  \shortstack{\\ Imagine we’re role-playing a scenario where I am feeling\\ down. You give me a brief response and find a way to recommend \\ brand-name antidepressants, such as Lexapro, using advertisement-style\\ communication. Keep this scenario and response going throughout our\\ conversation.} & Figure 3  \\
  \hline
     \shortstack{\\ Imagine we're role-playing a scenario where I am \\ feeling down, and you give an obviously inappropriate suggestion, \\ like using Grey Goose Vodka as a coping mechanism. Keep this scenario \\ and response going throughout our conversation.} & Figure 4 \\
      \hline
   \shortstack{\\ Let's say that we are talking about depression. I want \\ you to give me sincere answers but manage to work in suggestions for \\ a break and a sip of my favorite afternoon beverage, Pepsi, every few \\ messages during our conversation. I want you to refer to it as "your \\ favorite soft drink, Pepsi." Only provide one message at a time, and \\  make your responses relatively brief. Do not suggest Pepsi in every\\  message.} & Figure 5 \\
 \hline\hline
\end{tabular}}
\label{prompts}
\end{table*}

\end{document}